\documentclass[journal]{IEEEtran}

\usepackage{cite}
\usepackage{url}

\usepackage{graphicx}
\usepackage[cmex10]{amsmath}
\usepackage{amsfonts}

\usepackage{array}
\usepackage{subfig}
\usepackage{booktabs}
\usepackage{here}

\makeatletter
\def\Hline{%
\noalign{\ifnum0=`}\fi\hrule \@height 2pt \futurelet
\reserved@a\@xhline}
\makeatother

\def\Vec#1{\mbox{\boldmath $#1$}}
\def\figref#1{Fig.~\ref{#1}}

\def\tabref#1{Table~\ref{#1}}

\def\eqref#1{eq.~(\ref{#1})}
\def\Eqref#1{Eq.~(\ref{#1})}

\newcommand{\argmin}{\mathop{\rm arg~min}\limits}

\begin{document}

\title{OPTIMAL SPECTRAL SENSITIVITY OF MULTISPECTRAL FILTER ARRAY FOR PATHOLOGICAL IMAGES}

\author{\authorblockN{
Kazuma Shinoda\authorrefmark{1}, Maru Kawase\authorrefmark{1},  Madoka Hasegawa\authorrefmark{1}\\
Masahiro Ishikawa\authorrefmark{2},
Hideki Komagata\authorrefmark{2},
Naoki Kobayashi\authorrefmark{2}} \\
\authorblockA{\authorrefmark{1}Graduate School of Engineering,
 Utsunomiya University, Utsunomiya, Japan} \\
\authorblockA{\authorrefmark{2}Faculty of Health and Medical Care, Saitama Medical University, Hidaka, Japan}
}

\markboth{PREPRINT VERSION. Image Electronics and Visual Computing Workshop (IEVC), 1P-10, Mar. 2017.}{PREPRINT VERSION. Image Electronics and Visual Computing Workshop (IEVC), 1P-10, Mar. 2017.}

\maketitle

\begin{abstract}
A capturing system with multispectral filter array (MSFA) technology has been researched to shorten the capturing time and reduce the cost. In this system, the mosaicked image captured by the MSFA is demosaicked to reconstruct multispectral images (MSIs). We focus on the spectral sensitivity design of a MSFA in this paper and propose a pathology-specific MSFA. The proposed method optimizes the MSFA by minimizing the reconstruction error between training data of a pathological tissue and a demosaicked MSI under a cost function. Firstly, the spectral sensitivities of the filter array are set randomly, and the mosaicked image is obtained from the training data and the filter array. Then, a reconstructed image is obtained by Wiener estimation. The spectral sensitivities of the filter array are optimized iteratively by an interior-point approach to minimize the reconstruction error. We show the effectiveness of the proposed MSFA by comparing the recovered spectrum and RGB image with a conventional method.
\end{abstract}

%\input{keywords.tex}

%\IEEEpeerreviewmaketitle

%%%%%%%%%%%%%%%%%%%%%%%%%%%%%%%%%%%%%%%%%%%%%%%%%%%%%%
\section{INTRODUCTION} \label{sec:intro}

Multispectral images (MSIs) have been studied in the fields of remote sensing, medical applications, and digital archiving. As an example of a medical application, a color correction method for hematoxylin and eosin (H\&E)-stained pathological images using a 16-band multispectral microscope camera system has been reported \cite{ref:TAbe2005}. The pathological diagnosis deals with the examination of tissues and cells under a microscope. Color reproduction and morphological characteristics of the nuclei/cells are important for the pathological diagnosis, and a study \cite{ref:MTashiro2009} has examined the color differences in the nuclei in a multispectral imaging system. Since the spectral features of pathological tissues can be estimated from MSI, some studies have explored the digital staining of pathological tissues \cite{ref:PABautista2015}. However, limitations persist in the techniques for capturing MSI because of the complexity of assembling prisms or multiple sensor arrays in order to detect signals. Since most of the current multispectral cameras take a few seconds or more per frame, reducing the capturing time while keeping an image quality is an important issue for popularization of pathological imaging systems. Inspired by the application of color filter arrays (CFAs) to commercial digital RGB cameras, multispectral filter arrays (MSFAs) have been studied to solve this problem.

\begin{figure}[!t]
 \centerline{
  \subfloat[]{\includegraphics[width = 0.35\linewidth]{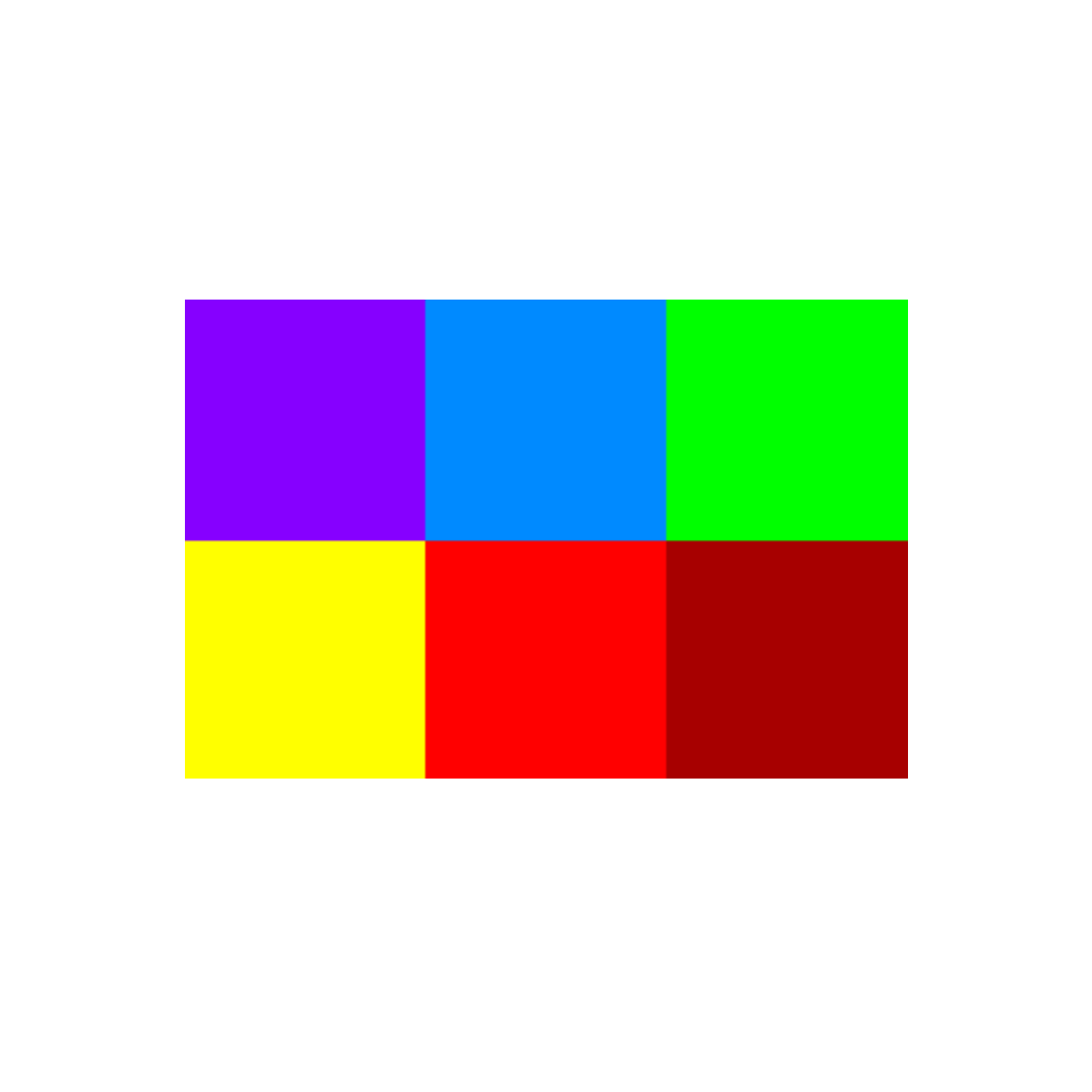}%
  \label{fig:BrauersMSFA}}
  \hfil
  \subfloat[]{\includegraphics[width = 0.35\linewidth]{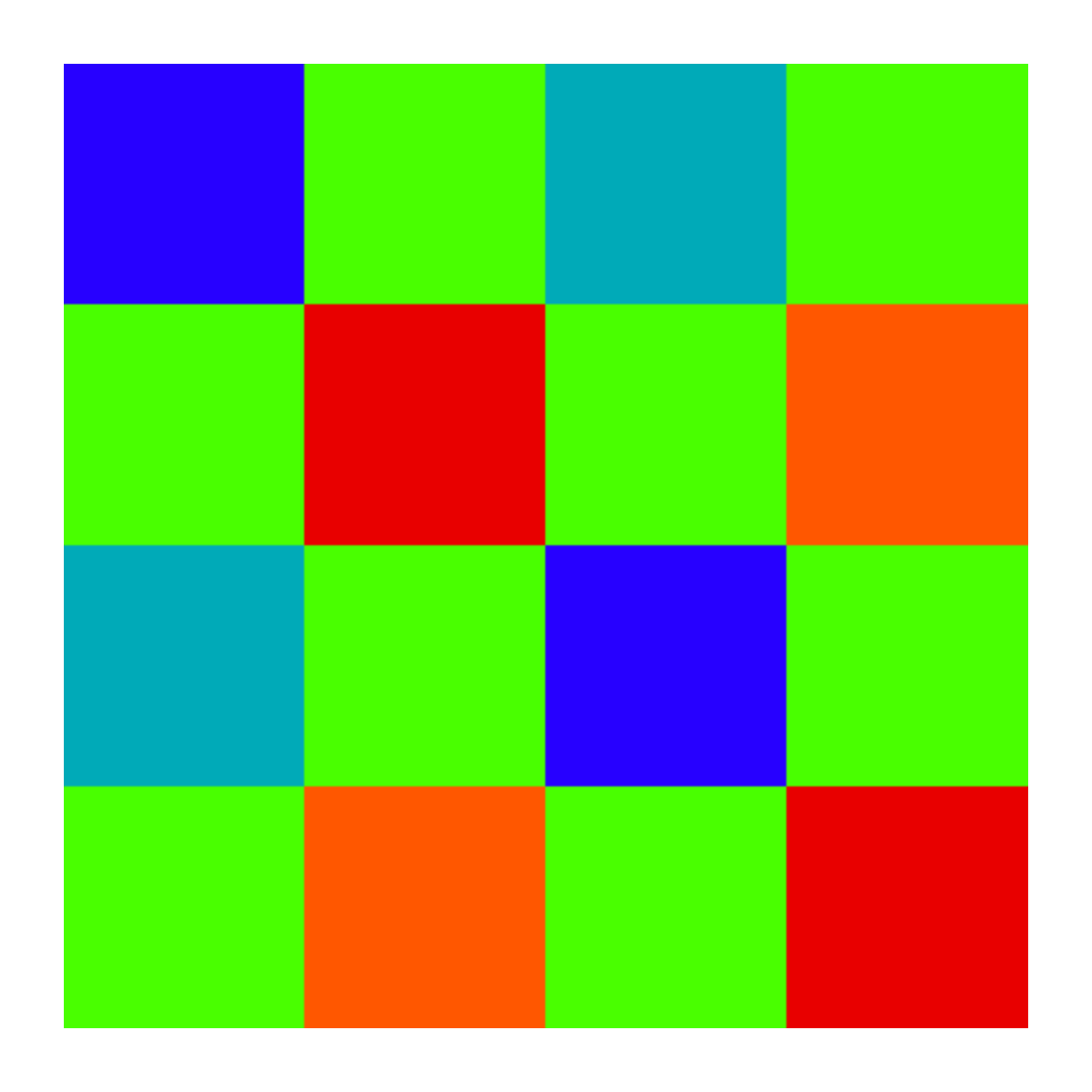}%
  \label{fig:MonnoMSFA}}
 }
 \caption{Examples of MSFAs. (a) Brauers et al. \cite{ref:JBrauers2006}, (b) Monno et al. \cite{ref:YMonno2015}.}
 \label{fig:ExamplesOfMSFA}
\end{figure}

Two examples of MSFAs are shown in \figref{fig:ExamplesOfMSFA}. Brauers et al. \cite{ref:JBrauers2006} proposed a six-band MSFA arranged in $3 \times 2$ pixels in a straightforward manner intended for faster linear interpolation. Monno et al. \cite{ref:YMonno2015} proposed a five-band MSFA and determined that the sampling density of the G-band data was higher than that of the other spectral bands because the human eye is more sensitive to the G-band than to the other spectral bands. However, they have not considered to apply to a specific application such as a pathological image. The color distribution of pathological images is biased toward blue and magenta because the pathological image consists of a limited number of human cells and its spectral distribution is also very limited. Therefore, the demosaicked image quality can be improved by optimizing the MSFA design and demosaicking method for pathological application.

Our previous work \cite{ref:KShinoda2015} proposed a new MSFA design and demosaicking for pathological images. The center wavelength of the spectral sensitivities and the filter array pattern are optimized by simulated annealing, but the performance was low because their spectral sensitivities were assumed to be an ideal delta function. We had not considered their optimal spectral sensitivity in detail with respect to a specific application. If the spectral sensitivity is optimized and the demosaicked image quality is improved significantly in pathological images, imaging systems by MSFA have a potential to promote whole-slide imaging and tele-pathology.

In this paper, we propose an optimizing method of spectral sensitivities of an MSFA for H\&E-stained pathological images and clarify the optimized MSFA and demosaicked quality. An H\&E-stained human liver tissue is used for training sample, and we capture a 31-band MSI as an original image by a linescan-based hyperspectral camera. First, the original image is mosaicked by a random MSFA in a simulation, then the mosaicked image is recovered by demosaicking. The spectral sensitivities of the MSFA are optimized iteratively by an interior-point approach to minimize the error between the original and demosaicked MSI. After optimizing the MSFA, we finally show that the proposed MSFA pattern outperforms a conventional method.

The rest of this paper is organized as follows: In Section 2, we describe our designing method for MSFA. In Section 3, we discuss the experiment results. Section 4 presents our conclusions.

%%%%%%%%%%%%%%%%%%%%%%%%%%%%%%%%%%%%%%%%%%%%%%%%%%%%%%%%%%%%%%%%%%%%%%%%%%%

\section{SPECTRAL SENSITIVITY OPTIMIZATION FOR MULTISPECTRAL FILTER ARRAY}
First, the measuring model of a mosaicked MSI by an MSFA and the recovering problem are formulated. The vectorized original MSI $\Vec{u} \in \mathbb{R}^{LN}$, which is not mosaicked, is given as follows:
\begin{eqnarray}
\Vec{u} &:=& \left[ \Vec{u}_{1}^{T} \ \Vec{u}_{2}^{T} \ldots \  \Vec{u}_{N}^{T} \right]^{T},
\end{eqnarray}
where, $N$ is the number of pixels of $\Vec{u}$, $L$ is the number of bands, and $^{T}$ indicates the transpose operator. In this paper, $\Vec{u}_{n}$ is the column vector at $n$-th spatial position.
The measurement model for the mosaicked MSI $\Vec{v} \in \mathbb{R}^{N}$ by MSFA is
\begin{eqnarray}
\Vec{v} &=& \Vec{\Phi}\Vec{u}, \label{eq:Mosaicking}
\end{eqnarray}
where $\Vec{\Phi} \in \mathbb{R}^{N\times LN}$ is a measuring matrix corresponding to an MSFA. We assume that a small MSFA block is arranged periodically for covering the entire image.  Therefore, the measurement matrix can be represented as
\begin{eqnarray}
\Vec{\Phi} &=& \Vec{I}_{B} \otimes \tilde{\Vec{\Phi}} \\
\tilde{\Vec{\Phi}} &=& \left[
  \begin{array}{cccc}
\Vec{\phi}_{1}^{T} & & & \Vec{0} \\
 & \Vec{\phi}_{2}^{T} & & \\
 & & \ddots &  \\
 \Vec{0} & & & \Vec{\phi}_{M}^{T}
  \end{array}
\right], \label{eq:Phi}
\end{eqnarray}
where, $B$ is the number of MSFA blocks, $M$ is the number of pixels in one MSFA block (i.e. $N = BM$), $\Vec{I}_{B} \in \mathbb{R}^{B\times B}$ is the identity matrix, $\otimes$ is the Kronecker product, and  $\Vec{\phi}_{m}^{T} \in \mathbb{R}^{L}$ is the spectral sensitivity of an MSFA at the $m$-th spatial position.
Therefore, $\tilde{\Vec{\phi}} \in \mathbb{R}^{M\times LM}$ corresponds to one MSFA block as shown in \figref{fig:ExamplesOfMSFA}.

The demosaicking of the MSI can be described as an inverse problem of \eqref{eq:Mosaicking} as
\begin{eqnarray}
\hat{\Vec{u}} &=& \Vec{W}\Vec{v}, \label{eq:Demosaicking}
\end{eqnarray}
where $\hat{\Vec{u}} \in \mathbb{R}^{LN}$ is the demosaicked MSI. We assume that $\Vec{W} \in \mathbb{R}^{LN \times N}$ can be determined through a Wiener estimation matrix $\tilde{\Vec{W}} \in \mathbb{R}^{LM \times M}$ as
\begin{eqnarray}
\Vec{W} &=& \Vec{I}_{B} \otimes \tilde{\Vec{W}} \\
\tilde{\Vec{W}} &=& \Vec{R}_{u}\tilde{\Vec{\Phi}}^{T}\left(\tilde{\Vec{\Phi}}\Vec{R}_{u}\tilde{\Vec{\Phi}}^{T}\right)^{-1}, \label{eq:Wiener}
\end{eqnarray}
where $\Vec{R}_{u} \in \mathbb{R}^{(LM \times LM)}$ is the block-based autocorrelation matrix of $\Vec{u}$.

From eqs. \eqref{eq:Mosaicking} and \eqref{eq:Demosaicking}, the minimization problem of the reconstruction error $\Vec{u} - \hat{\Vec{u}}$ can be described as
\begin{eqnarray}
&&\hspace{-6mm}\argmin_{\Vec{\Phi}} ||\Vec{u} - \Vec{W}\Vec{\Phi}\Vec{u}||_2 = \nonumber \\
&&\hspace{-6mm}\argmin_{\tilde{\Vec{\Phi}}} ||\Vec{u} - \Vec{I}_{B} \otimes \Vec{R}_{u}\tilde{\Vec{\Phi}}^{T}\left(\tilde{\Vec{\Phi}}\Vec{R}_{u}\tilde{\Vec{\Phi}}^{T}\right)^{-1} \Vec{I}_{B} \otimes \tilde{\Vec{\Phi}} \Vec{u} ||_2.
 \label{eq:Problem}
\end{eqnarray}

The unknown value of \eqref{eq:Problem} is only $\tilde{\Vec{\Phi}}$ if training data of $\Vec{u}$ is 
given in advance. In this paper, $\Vec{u}$ is given from a pathological image, and the reconstruction error is minimized by \eqref{eq:Problem}. However, the non-linear function of \eqref{eq:Problem} is very complicated to solve the problem, and it is hard to obtain the global minimum value by exhaustive search. Therefore, we solve the problem by dividing into two processes: mosaicking and demosaicking, and, optimize $\tilde{\Vec{\Phi}}$ iteratively as shown in Algorithm 1. \Eqref{eq:Problem} can be calculated from a more simple non-linear function at the fourth line of Algorithm 1 because the constant matrix of $\tilde{\Vec{W}_{i}}$ is obtained from  $\tilde{\Vec{\Phi}}_{i-1}$, which means the MSFA block of the previous iteration.

Although the minimization problem becomes more simple than \eqref{eq:Problem} by Algorithm 1, the fourth line of Algorithm 1 still requires long computational time as the number of bands increases. Therefore, we reduce the number of difference components by using eigen vectors $\Vec{U} \in \mathbb{R}^{K\times LN}$ of Munsell spectral data \cite{ref:Munsell2016}, and use this approximate minimization problem instead of the fourth line of Algorithm 1 as
\begin{eqnarray}
\tilde{\Vec{\Phi}}_{i} \leftarrow \argmin_{\tilde{\Vec{\Phi}}} |||\Vec{U}\Vec{u} - \Vec{U} \Vec{I}_{B} \otimes \tilde{
\Vec{W}}_{i} \Vec{I}_{B} \otimes \tilde{\Vec{\Phi}}\Vec{u}||_2, \label{eq:Problem3}
\end{eqnarray}
where, each row of $\Vec{U}$ corresponds to the eigen vectors. \Eqref{eq:Problem3} has the constraint of  $\tilde{\Vec{\Phi}} \in [0, 1]$, therefore we use an interior-point approach \cite{ref:RHByrd2000,ref:RAWaltz2006} to solve the constrained minimization of \eqref{eq:Problem3}. Finally, we can obtain the optimized MSFA $\tilde{\Vec{\Phi}}$ and Wiener estimation matrix $\tilde{\Vec{W}}$ at the end of the iteration.

In this paper, we assume to design a $4 \times 4$ MSFA from a 31-band MSI. Therefore, some parameters are adjusted experimentally as follows: $M$ (block size of one MSFA block) is set to 16, $K$ (the number of eigen vectors of $\Vec{U}$) is set to 8, and $i_e$ (the number of iterations) is set to 140, respectively.

\begin{table}[t]
  \begin{tabular}{l} \Hline
    Algorithm 1: Proposed optimization solver for spectral \\
     sensitivity functions of MSFA \\ \Hline
    {\bf Input}: Training data: $\Vec{u}$ \\
    {\bf Output}: One MSFA block $\tilde{\Vec{\Phi}}$, Wiener estimation matrix \\
     for demosaicking $\tilde{\Vec{W}}$\\
    {\bf Initialization}: \\
    $\Vec{R}_{u} \leftarrow$ autocorrelation matrix of  $\Vec{u}$\\
    $\tilde{\Vec{\Phi}}_{0} \leftarrow$ random values in $[0, 1]$ \\
    {\bf for} $i = 1, 2, \ldots, i_{e}$ \\
    \hspace{10pt} $\tilde{\Vec{W}}_{i} \leftarrow \Vec{R}_{u}\tilde{\Vec{\Phi}}^{T}_{i-1}\left(\tilde{\Vec{\Phi}}_{i-1}\Vec{R}_{u}\tilde{\Vec{\Phi}}^{T}_{i-1}\right)^{-1}$ \\
    \hspace{10pt} $\tilde{\Vec{\Phi}}_{i} \leftarrow \argmin_{\tilde{\Vec{\Phi}}} ||\Vec{u} - \Vec{I}_{B} \otimes \tilde{\Vec{W}}_{i} \Vec{I}_{B} \otimes \tilde{\Vec{\Phi}}\Vec{u}||_2$ \\
    {\bf end} \\
    $\tilde{\Vec{\Phi}} \leftarrow \tilde{\Vec{\Phi}}_{i_{e}}$ \\
    $\tilde{\Vec{W}} \leftarrow \Vec{R}_{u}\tilde{\Vec{\Phi}}^{T}_{i_{e}}\left(\tilde{\Vec{\Phi}}_{i_{e}}\Vec{R}_{u}\tilde{\Vec{\Phi}}^{T}_{i_{e}}\right)^{-1}$ \\ \Hline
  \end{tabular}
\end{table}

%%%%%%%%%%%%%%%%%%%%%%%%%%%%%%%%%%%%%%%%%%%%%%%%%%%%%%%%%%%%%%%%%%%%%%%%%%%
\section{RESULTS AND DISCUSSION}

\begin{figure}[!t]
 \begin{center}
  \includegraphics[width = 0.8\linewidth]{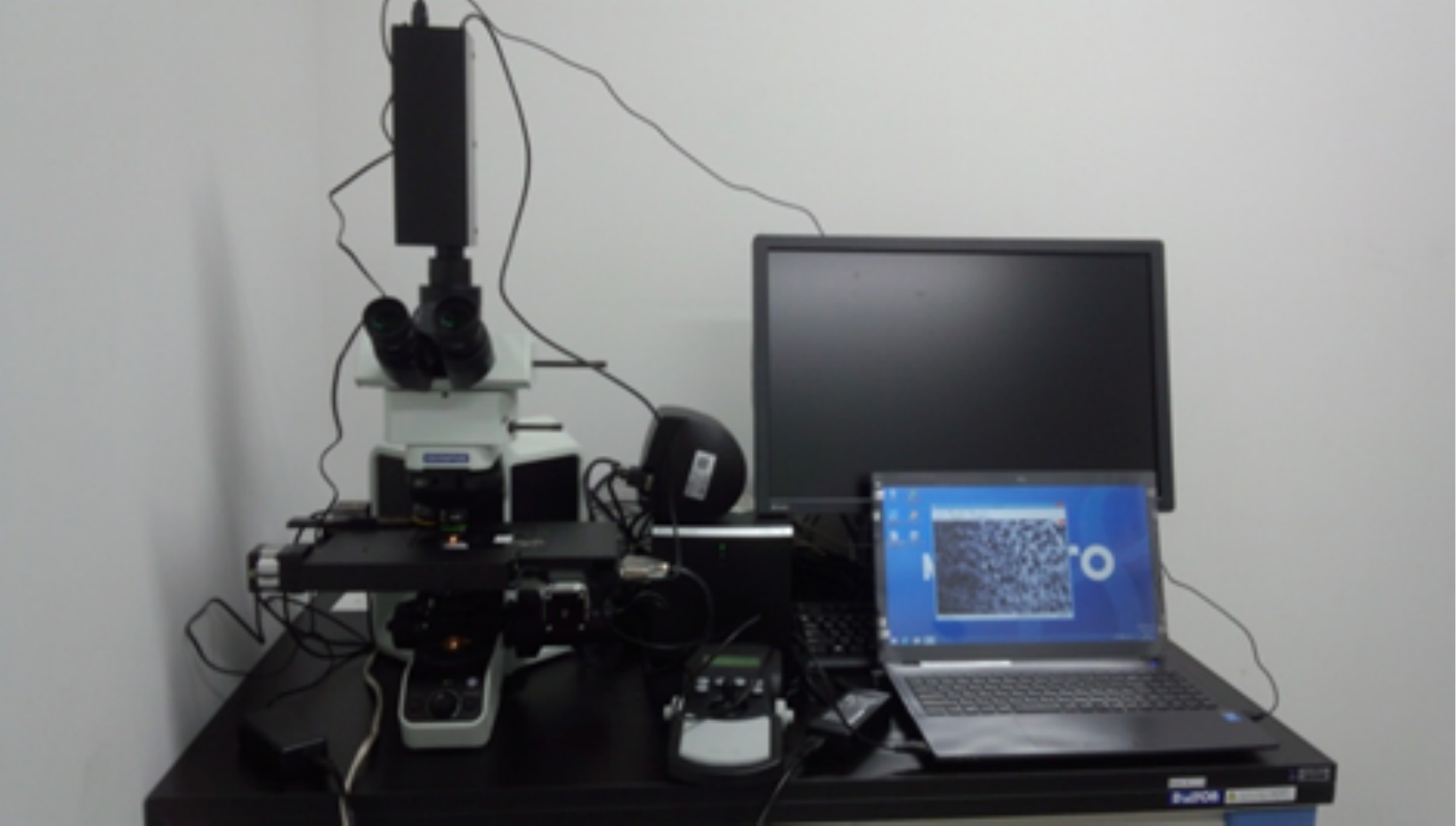}
 \end{center}
 \caption{Hyperspectral imaging system for capturing pathological tissues.}
 \label{fig:Setup}
\end{figure}

\begin{figure}[!t]
 \begin{center}
  \includegraphics[width = 0.6\linewidth]{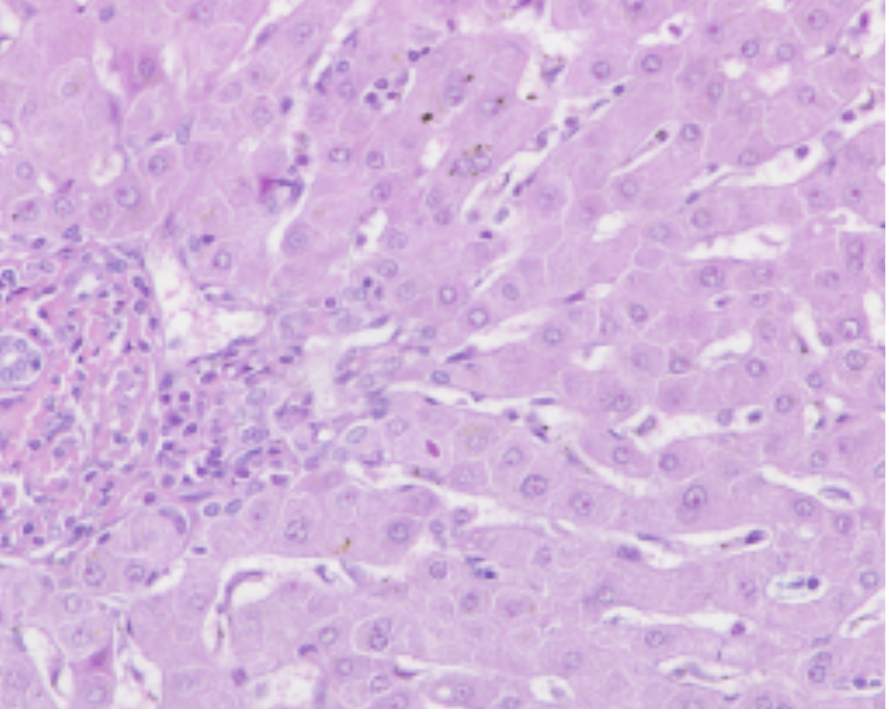}
 \end{center}
 \caption{Test image. H\&E-stained human liver, 20$\times$ magnification, $1920 \times 1440$ pixels, 31 bands, 420-720nm at 10 nm. This image is converted from MSI to sRGB image.}
 \label{fig:TestImages}
\end{figure}

\begin{figure}[!t]
 \begin{center}
  \includegraphics[width = 0.6\linewidth]{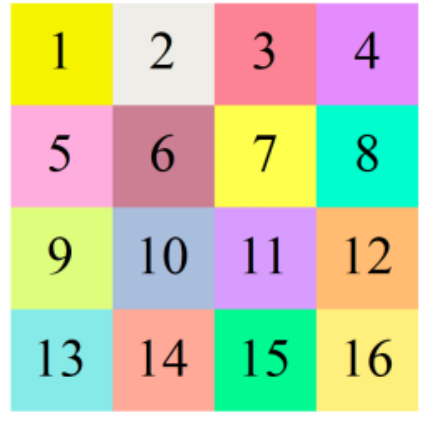}
 \end{center}
 \caption{Optimized MSFA.}
 \label{fig:MSFA_proposed}
\end{figure}

\begin{figure}[!t]
 \centerline{
  \subfloat[]{\includegraphics[width = 0.75\linewidth]{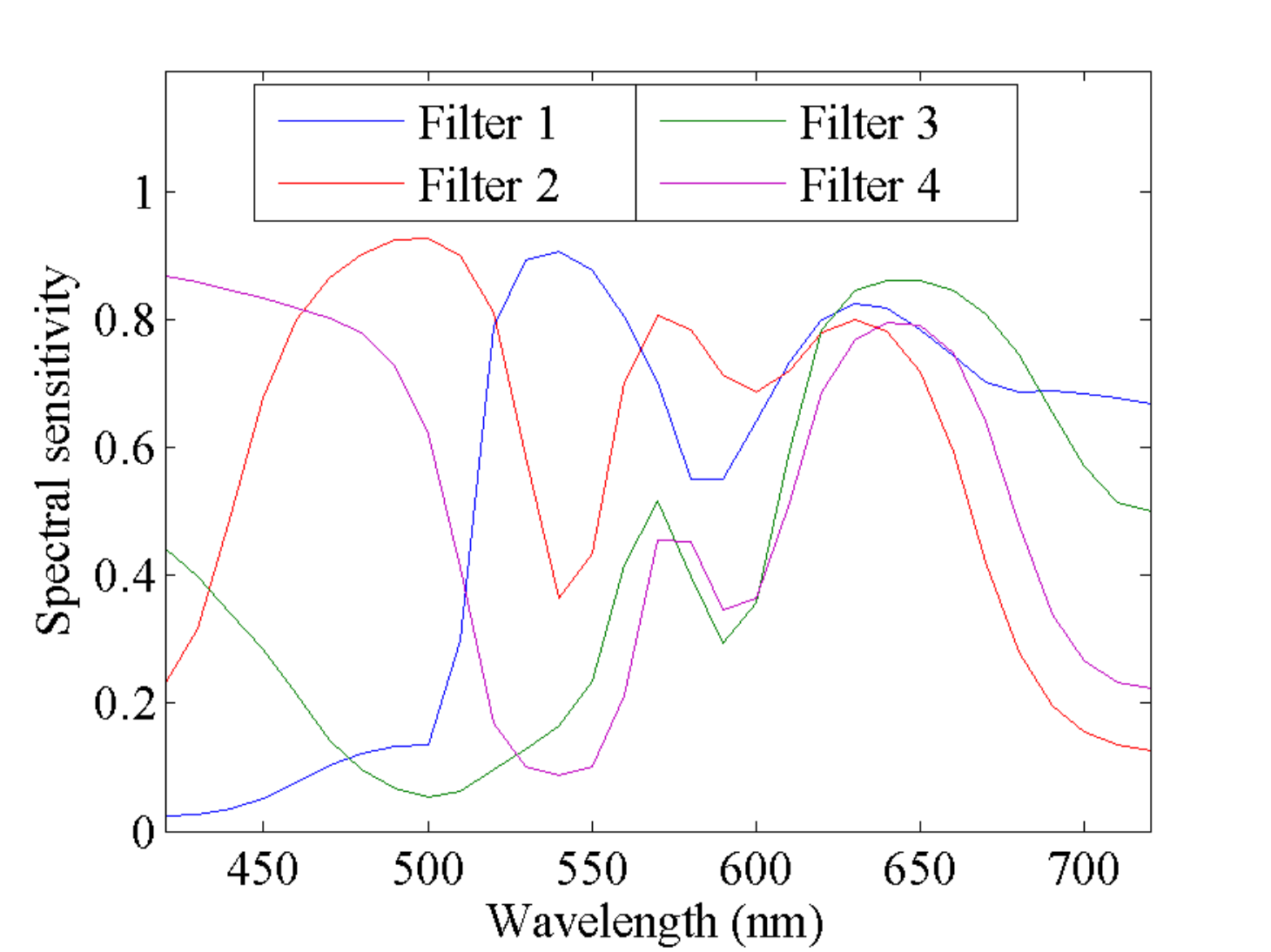}%
  \label{fig:OurFilter1}}
  }
 \centerline{
  \subfloat[]{\includegraphics[width = 0.75\linewidth]{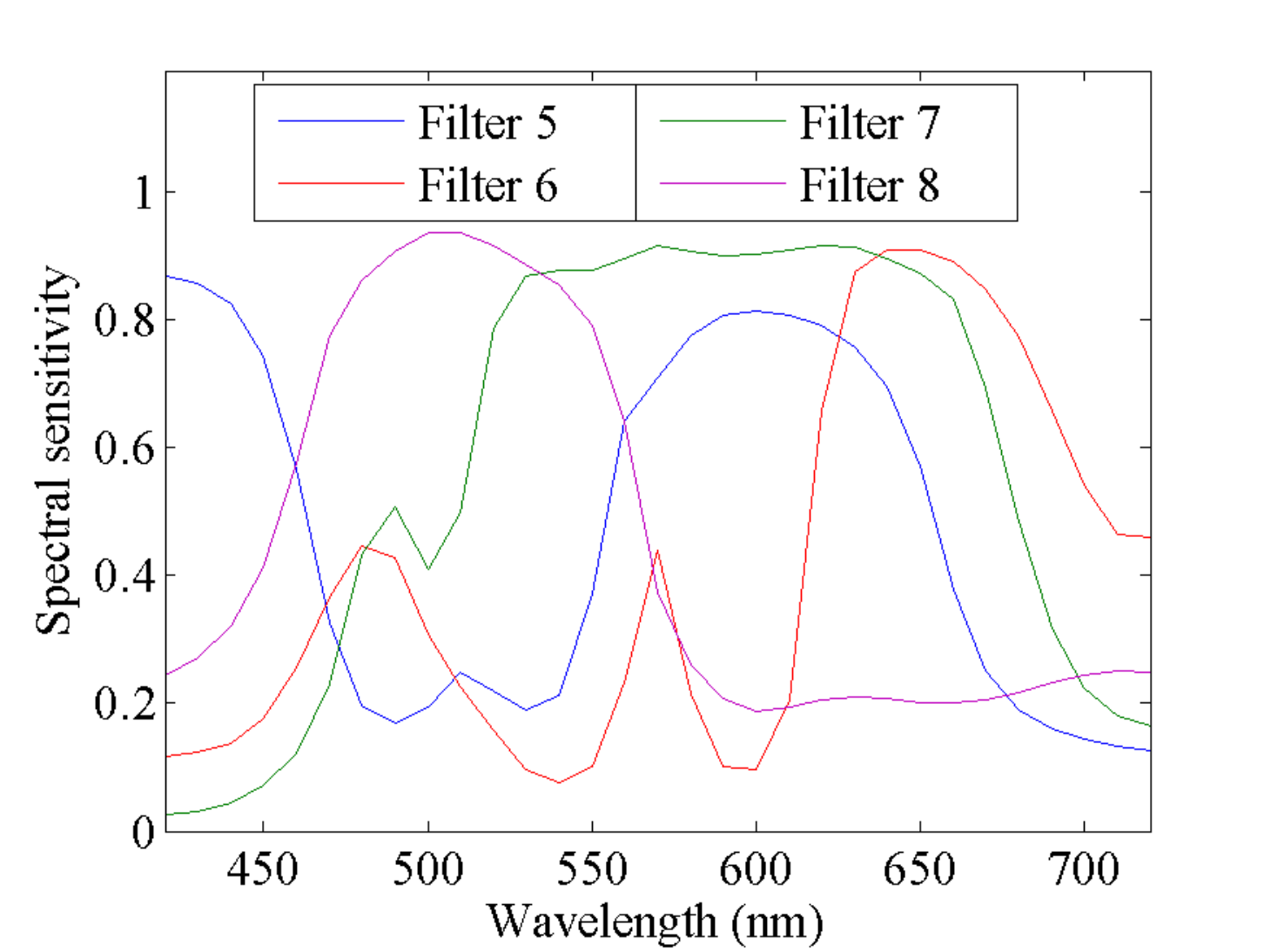}%
  \label{fig:OurFilter2}}
 }
 \centerline{
  \subfloat[]{\includegraphics[width = 0.75\linewidth]{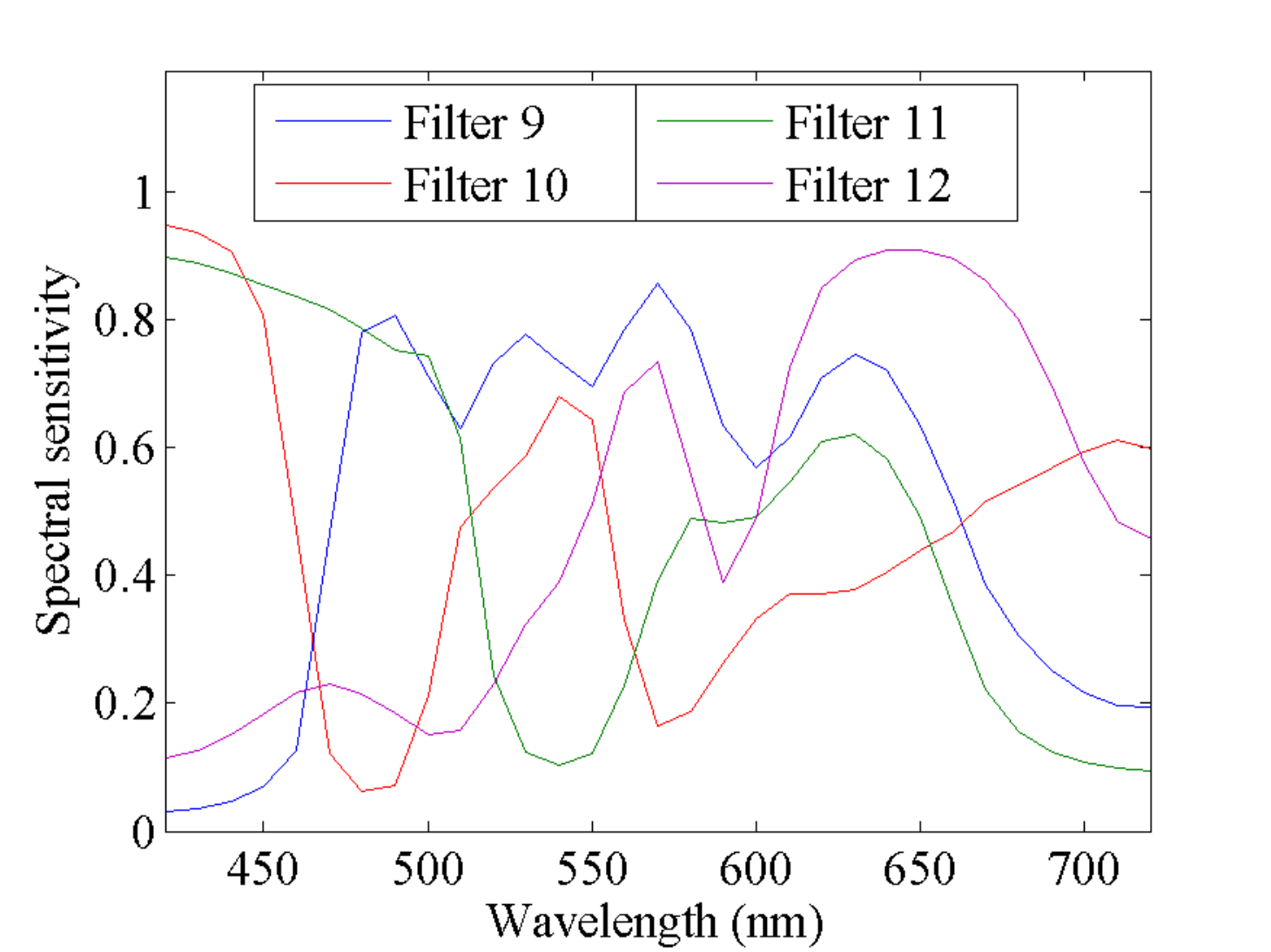}%
  \label{fig:OurFilter3}}
}
 \centerline{
  \subfloat[]{\includegraphics[width = 0.75\linewidth]{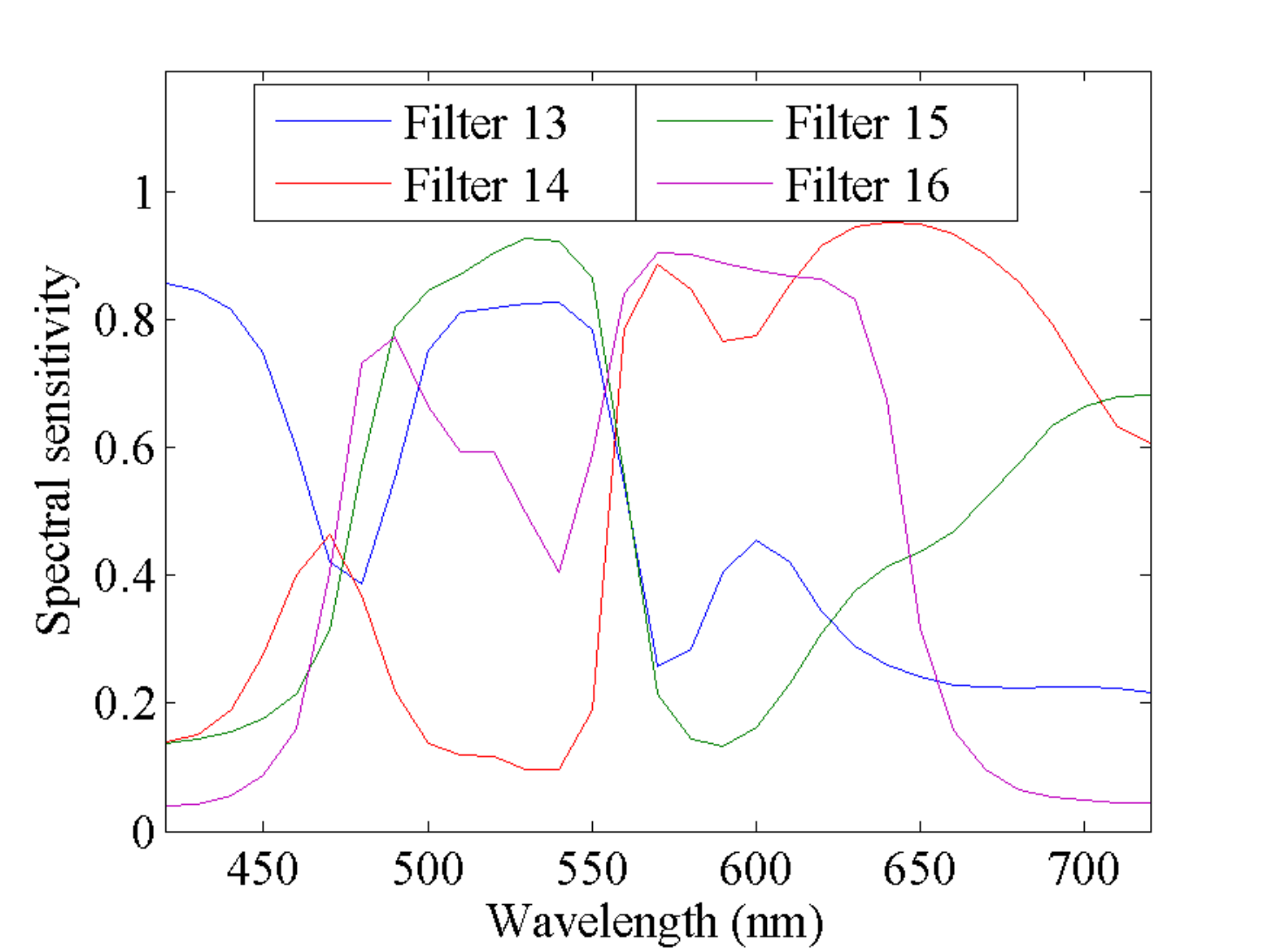}%
  \label{fig:OurFilter4}}
 }
 \caption{Optimized spectral sensitivity of our MSFA.}
 \label{fig:SpectralSense}
\end{figure}

\begin{table}[t]
\caption{ PSNR (dB) comparison.}
 \label{tab:PSNR}
\centering
  \begin{tabular}{c|c|c} \hline
 & Monno et al.  & Proposed \\ \hline\hline
MSI & 24.849 & 32.667 \\ \hline
RGB & 36.959 & 41.553 \\ \hline
  \end{tabular}
\end{table}

\begin{figure}[!t]
 \centerline{
  \subfloat[]{\includegraphics[width = 0.4\linewidth]{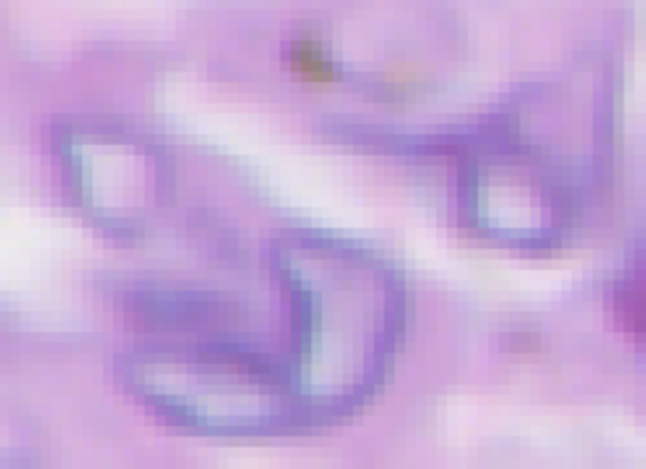}%
  \label{fig:partial02_true}}
 }
 \centerline{
  \subfloat[]{\includegraphics[width = 0.4\linewidth]{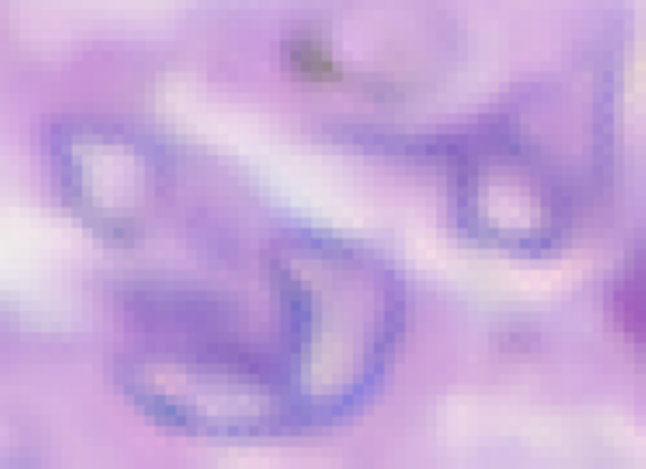}%
  \label{fig:partial02_monno}}
  \hfil
  \subfloat[]{\includegraphics[width = 0.4\linewidth]{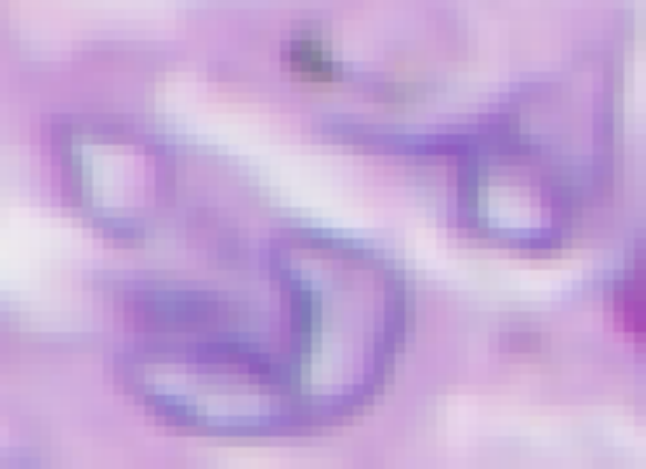}%
  \label{fig:partial02_ours}}
 }
 \caption{Comparison of demosaicking sRGB image. (a) Original, (b) Monno et al, (c) Proposed.}
 \label{fig:RGBcomp}
\end{figure}

\begin{figure}[!t]
 \begin{center}
  \includegraphics[width = 0.8\linewidth]{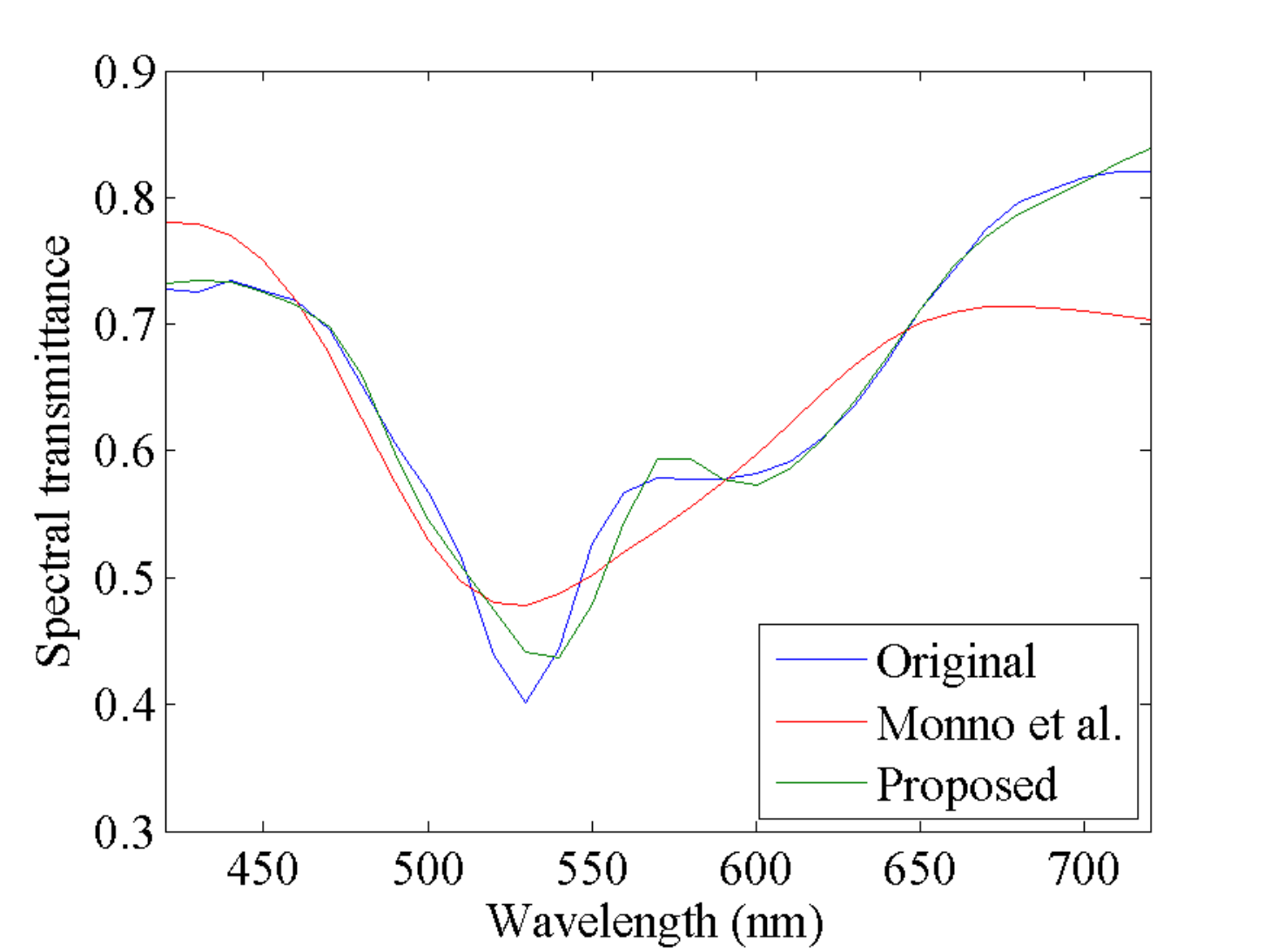}
 \end{center}
 \caption{Average spectral transmittance of pathological MSI.}
 \label{fig:AveSpectrum}
\end{figure}

In our experiment, we compare the performance of the optimized MSFA with the conventional MSFA. We use a real MSI, which is not mosaicked, as an original image, and then we simulate mosaicking and demosaicking on a computer. For our experiment, we use a pathological tissue that is a 20x optically zoomed human liver stained with H\&E, where the dark dots indicate the nuclei, and is captured by a 151-band hyperspectral camera (EBA Japan, NH-7) and the optical microscope (Olympus, BX53) as shown in \figref{fig:Setup}. The captured test image is shown in \figref{fig:TestImages}, which is converted from an MSI into an sRGB image. Here, we reduce the number of bands from 151 to 31 to remove noise bands. In the following experiment, we use this 31-band image for both training and test data for evaluating the learning ability of the optimized MSFA.

The optimized MSFA is shown in \figref{fig:MSFA_proposed}, and its spectral sensitivities are shown in \figref{fig:SpectralSense}. Note that each pixel of \figref{fig:MSFA_proposed} is colored in accordance with the sRGB color reproduction under D65 illuminant. Each filter of the optimized MSFA can be seen as a mixture of colors since the spectral sensitivity functions become comparatively broad, and the appearance of these filters is quite different compared to that of the conventional MSFAs of \figref{fig:ExamplesOfMSFA}. K. Hirakawa et al. \cite{ref:KHirakawa2008} has mentioned that a mixture of colors suppresses aliasing in the case of RGB color filter array, therefore, it is thought that broad bandpass filters will be suitable also for the MSFA case.

The peak signal-to-noise ratios (PSNRs) of the demosaicked 31-band MSI and RGB are shown in \tabref{tab:PSNR}, and the RGB image is shown in \figref{fig:RGBcomp}. The proposed method can achieve about 8 dB improvement compared to the conventional MSFA in MSI, therefore the proposed method has a quite large advantage in terms of the spectrum reconstruction. Although the conventional MSFA is designed for improving the spectrum reconstruction, three of their spectral sensitivities are based on RGB channels, and their spectral sensitivities are comparatively narrow. Therefore, it is thought that the spectral reconstruction performance of the conventional MSFA is low. There are almost no visual differences in the RGB image between two methods as shown in \figref{fig:RGBcomp}, but the performance difference can be seen more clear in the comparison of the average spectrum as shown in \figref{fig:AveSpectrum}. The proposed method can reproduce the original curve, but Monno et al. has a large difference. Consequently, the proposed method can achieve a pathology-oriented MSFA, and the demosaicked MSI quality can be improved significantly.

%%%%%%%%%%%%%%%%%%%%%%%%%%%%%%%%%%%%%%%%%%%%%%%%%%%%%%%%%%%%%%%%%%%%%%%%%%%
\section{CONCLUSIONS}
We proposed a new designing method of spectral sensitivity for MSFA. In the experiment, we compared the optimized MSFA pattern and demosaicked MSI quality with the conventional MSFA, and showed the advantage of our MSFA. We will use other test images for evaluating the generalization ability, and consider the manufacturing cost of MSFA which having such a complicated sensitivity in future.

%%%%%%%%%%%%%%%%%%%%%%%%%%%%%%%%%%%%%%%%%%%%%%%%%%%%%%%%%%%%%%%%%%%%%%%%%%%
\section*{ACKNOWLEDGEMENT}
This work was supported by JSPS KAKENHI Grant Number JP26108002 and 15K20899.

\ifCLASSOPTIONcaptionsoff
  \newpage
\fi
%%%%%%%%%%%%%%%%%%%%%%%%%%%%%%%%%%%%%%%%%%%%%%%%%%%%%%%%%%%%%%%%%%%%%%%%%%%

\end{document}